\documentclass[aps,prl,reprint]{revtex4-1}
\usepackage{graphicx}
\usepackage{amsmath,amsmath}
\usepackage{stix}
\usepackage{amssymb}
\usepackage{setspace}
\usepackage{array}
\usepackage{multirow}
\usepackage{url}
\usepackage{color,soul}
\usepackage{float}
\usepackage{setspace}
\usepackage{lipsum}
\usepackage[pagewise]{lineno}
\usepackage{comment}
\usepackage{gensymb}
\usepackage{titlesec}
\usepackage{hyperref}
\usepackage{relsize}
\usepackage{textcomp}
\usepackage{xhfill}
\usepackage[width=17.00cm, height=23.00cm]{geometry}

    \newcommand{\sone}{\textcolor[RGB]{191,0,191}{$\blacktriangle$}}
    \newcommand{\sfive}{\textcolor[RGB]{0,191,191}{$\mdblkdiamond$}}
    \newcommand{\snine}{\textcolor[RGB]{0,255,0}{$\blacktriangleleft$}}
    \newcommand{\RB}{\textcolor[RGB]{0,0,0}{$\triangledown$}}
    \newcommand{\IH}{\textcolor[RGB]{0,0,255}{$\circ$}}
    \newcommand{\hlone}{\textcolor[RGB]{255,0,0}{$\blacksquare$}}
    
    \newcommand{\sseven}{\textcolor[RGB]{0,128,0}{$\star$}}

\newcommand{\vecu}{\mathbf{u}}

\begin{document}
      
\title{Transition between boundary-limited and mixing-length scalings\\of turbulent transport in internally heated convection}

\author{Sina Kazemi}
\affiliation{Department of Mechanical Engineering, University of Houston, Houston, TX 77040, USA}
\author{David Goluskin}
\affiliation{Department of Mathematics and Statistics
University of Victoria, Victoria, BC V8P 5C2}
\author{Rodolfo Ostilla-M\'onico}
\affiliation{Department of Mechanical Engineering, University of Houston, Houston, TX 77040, USA}
\affiliation{Escuela Superior de Ingener\'ia, Universidad de C\'adiz, C\'adiz, Spain}
\date{\today}

\begin{abstract}
Heat transport in turbulent thermal convection increases with the thermal forcing, but in almost all studies the rate of this increase is slower than it would be if transport became independent of the molecular diffusivities---the heat transport scaling exponent is smaller than the mixing-length (or `ultimate') value of $1/2$. This is due to thermal boundary layers that throttle heat transport in configurations driven either by thermal boundary conditions or by internal heating, giving a scaling exponent close to the boundary-limited (or `classical) value of $1/3$. With net-zero internal heating and cooling in different regions, the larger mixing-length exponent can be attained because heat need not cross a boundary. We report numerical simulations in which heating and cooling are unequal. As heating and cooling rates are made closer, the scaling exponent of heat transport varies from its boundary-limited value to its mixing-length value.
\end{abstract}

\maketitle	
	 

\noindent 
Thermally driven turbulence is prevalent in natural phenomena and technological applications. Research on its fundamental physics has focused most on the canonical configuration of Rayleigh--B\'enard convection (RBC), where flow in a fluid layer is driven by a temperature gradient maintained between the top and bottom boundaries. However, many natural and technological systems are not driven solely by heat fluxes across the fluid's boundaries---they are driven partly or entirely by volumetric heat sources and/or sinks. Such internally heated convection (IHC) can be driven by various mechanisms. The Earth's mantle is heated by radioactive decay 
\citep{bercovici1989three,schubert2001mantle,houseman1988dependence,limare2015microwave}, other planets and their satellites are heated by tidal forces \citep{peale1979melting,moore2003tidal,sotin2002europa}, large stars have convective cores driven by thermonuclear reactions  \citep{kippenhahn1990stellar,spiegel1971convection}, and planetary atmospheres absorb solar radiation \citep{tritton1975internally,phillips1998geological}. In technological applications, fluids may be internally heated or cooled by chemical reactions \citep{shen2016thermal,ponce2015using}. Other cooling mechanisms include radiation and phase change, and heating and cooling mechanisms can coexist.

Simple configurations have served as fundamental models for IHC \citep{goluskin2016internally}, much as RBC is a fundamental model for convection driven by boundary heat fluxes. Two IHC models of uniformly heated fluid layers, differing only in their boundary conditions, have received the most attention, although far less than RBC. The first configuration has an insulating bottom and an isothermal top, so all heat flux is outward across the top. It has been studied experimentally \citep{tritton1967convection,schwiderski1971convection,takahashi2010experimental,lee2007boundary,tasaka2005experimental} and computationally \citep{mckenzie1974convection,thirlby1970convection,tveitereid1976convection,schubert1993steady,farouk1988turbulent,emara1980numerical}, but no three-dimensional (3D) direct numerical simulations (DNS) of the turbulent regime have been reported previously. The second configuration has isothermal top and bottom boundaries at equal temperatures, so heat flux is outward across both boundaries. It too has been studied experimentally \citep{kulacki1972thermal,jahn1974free,ralph1977experiments,lee2007boundary,peckover1974convective} and computationally \cite{jahn1974free,straus1976penetrative,tveitereid1978thermal,emara1980numerical,grotzbach1989turbulent,worner1997direct,goluskin2012convection}, including one modern DNS study in 3D \cite{goluskin2016penetrative}. In both of these models, certain bulk quantities connected to turbulent heat transport show strong similarities to RBC.

Heat transport in RBC is captured by the dimensionless Nusselt number ($Nu$), which is the ratio of total upward heat transport to that by conduction alone. Of particular interest is the dependence of $Nu$ on the dimensionless Rayleigh number ($Ra$), which is proportional to the dimensional temperature difference across the layer. The asymptotic scaling of $Nu$ as $Ra\to\infty$ remains unsettled, with the main predictions being $Nu\sim Ra^{1/3}$ or $Nu\sim Ra^{1/2}$; see Ref.\ \cite{doering2020turning} for an overview. Although RBC experiments and simulations have reached $Ra$ values beyond $10^{15}$, it is unclear whether such $Ra$ are large enough to produce asymptotic scaling of $Nu$. Many experiments show exponents close to $1/3$, and some show signs of a transition near the upper end of their $Ra$ ranges, but none have shown an exponent close to $1/2$ for ordinary RBC. The `mixing-length' scaling $Ra^{1/2}$ (often called `ultimate' scaling in the context of RBC)  is predicted by so-called mixing-length arguments \cite{spiegel1963generalization}; it occurs if and only if dimensional transport becomes independent of molecular diffusivities.  The `boundary-limited' scaling $Ra^{1/3}$ (often called `classical' scaling in RBC) has been attributed to mechanisms that prevent the thermal boundary layers from getting thin enough to conduct heat any faster than this \cite{howard1966convection}.

Boundary-limited and mixing-length scalings have meaning for heat transport in IHC models also. In IHC there are various reasonable ways to define a Nusselt number that give different values, whereas in RBC these definitions coincide \citep{goluskin2016internally}. Below we describe ways to define $Nu$ and $Ra$ for IHC such that, as in RBC, the scalings $Nu\sim Ra^{1/3}$ and $Nu\sim Ra^{1/2}$ reflect boundary-limited and mixing-length transport, respectively. For the two IHC models described above in which heating is uniform throughout the fluid, all past experiments and simulations have displayed boundary-limited scalings of heat transport, much like RBC. This is unsurprising since average outward heat flux across the boundaries must equal the internal generation, and all heat crossing a boundary must first traverse a boundary layer, as with all heat flowing into or out of the RBC domain. In systems with both internal heating and cooling, however, transport from cooled regions to heated regions need not cross a boundary layer, so scalings might not be boundary-limited.

Recent experiments and simulations \cite{lepot2018radiative,bouillaut2019transition,miquel2020} have revealed that mixing-length scaling of heat transport is indeed possible in convection that is both heated and cooled internally. In these studies the fluid is insulated on all sides and subject to heating/cooling proportional to $ae^{-z/\ell}-
\beta$, where $z$ measures distance above the bottom boundary, the other quantities are parameters, and $
\beta$ is chosen so that net heating/cooling is zero. There is heating below the height $z=-\ell\log (\beta/a)$ and cooling above it. The scale $\ell$ over which heating decreases exponentially with height is chosen small enough that the heating region at the bottom is thinner than the cooling region at the top. 
When $\ell$ is very small, the heating region concentrates at the bottom boundary, a boundary layer forms, and transport scaling is boundary-limited. When $\ell$ is large enough for the heating region to be thicker than such a boundary layer, transport between the heating and cooling regions displays mixing-length scaling. This is consistent with the observation of mixing-length scaling in other modifications of RBC that circumvent the boundary layers, for instance using boundary roughness \cite{toppaladoddi2017roughness} or by decoupling the thermal and velocity boundary layers~\cite{zou2021realizing}.

The configuration with net-zero heating/cooling in which Refs.\ \cite{lepot2018radiative,bouillaut2019transition,miquel2020} observe mixing-length scaling is an extreme case where no heat crosses the boundaries. At the opposite extreme are the RBC model and the IHC models with no cooling, where all transported heat must cross a boundary, and where boundary-limited scalings are seen at all accessible $Ra$ values. In between these extremes, however, lies any convective system with unequal rates of heating and cooling. When heating exceeds cooling, say, some internally produced heat is transported only to a cooling region, but some must escape across a boundary layer. Our primary aim here is to determine whether such systems display boundary-limited scaling, mixing-length scaling, or something in between.

In the present work we carry out 3D DNS of turbulent convection driven by an internal heating/cooling profile of the form $ae^{-z/\ell}-\beta$, but unlike Refs.\ \cite{lepot2018radiative,bouillaut2019transition,miquel2020} net heating is permitted. The bottom is insulating, but the top is made isothermal so that heat can escape. Without cooling ($
\beta=0$) we find boundary-limited scaling as expected, and we verify that spatially uniform heating gives similar results. With $\beta$ chosen to give net-zero heating/cooling, so that our configuration is like Refs.\ \cite{lepot2018radiative,bouillaut2019transition,miquel2020} but with an isothermal top, we find mixing-length scaling as expected. We then carry out DNS over a range of $Ra$ at various smaller $\beta$ values, each of which corresponds to a different fraction of internally produced heat that must cross the top boundary. Altering this fraction creates a transition between boundary-limited and mixing-length scalings of convective transport. In terms of the $Nu$ and $Ra$ defined below for IHC, we observe approximate power laws $Nu\sim Ra^\alpha$. As the rate of cooling is raised from zero until it equals the rate of heating, the exponent $\alpha$ increases from about 0.3 to $0.5$; the analogue in RBC is a transition from classical to ultimate scaling.


Convection with internal sources or sinks can be modeled with the Oberbeck--Boussinesq approximation. The layer height $d$ and thermal diffusivity $\kappa$ define length and time scales of $d$ and $d^2/\kappa$, respectively. If the heating/cooling profile is proportional to a rate $Q$ with units of temperature per time, $d^2Q/\kappa$ is a temperature scale. With these scales, the dimensionless equations for velocity $\vecu(\mathbf x,t)$, pressure $p(\mathbf x,t)$, and temperature $T(\mathbf x,t)$ are~\cite{goluskin2016internally}
\begin{subequations}
\label{eq: bouss}
\begin{align}
\nabla \cdot \vecu&=0, \label{eq: inc} \\
\partial_t \vecu+\vecu\cdot \nabla \vecu &=-\nabla p+Pr {{\nabla }^{2}}\vecu+Pr R\,T\mathbf{\hat{z}}, \label{eq: u eq} \\
\partial_t T+\vecu\cdot \nabla T&={{\nabla }^{2}}T+H(z), \label{eq: T eq}
\end{align}
\end{subequations}
where $z$ is the vertical component of the spatial coordinate $\mathbf x$, and the dimensional heating/cooling profile is $QH(dz)$. The Prandtl number is $Pr=\nu/\kappa$, where $\nu$ is kinematic viscosity. The other dimensionless control parameter is $R={g\alpha {{d}^{5}}Q}/{{\kappa }^{2}\nu}$, where $g$ is gravitational acceleration in the $-\mathbf{\hat z}$ direction and $\alpha$ is the linear coefficient of thermal expansion. This $R$ differs from the Rayleigh number $Ra$ of RBC in that the temperature scale is $d^2Q/\kappa$ rather than being the dimensional temperature difference $\Delta$ between the boundaries, meaning that $Ra=R\Delta/(d^2Q/\kappa)$. Here $Ra$ is a diagnostic quantity rather than a control parameter since $\Delta$ is dynamically determined.

The dimensionless domain is periodic in both horizontal directions with a period of $\Gamma$, and its vertical ($z$) extent is $[0,1]$. At the top ($z=1$) and bottom ($z=0$) boundaries we enforce no-slip conditions by $\vecu=\mathbf 0$. The top boundary is isothermal, and we call its temperature $T=0$. The bottom boundary is insulating, so $\partial_zT=0$ there.

Vertical heat transport has convective and conductive contributions, which are proportional to the dimensionless quantities $\mathbf{\hat{z}}\cdot \vecu\hspace{1pt}T$ and $-\partial_zT$, respectively. The mean \emph{total} transport is $\langle \mathbf{\hat{z}}\cdot \vecu\hspace{1pt}T-\partial_zT\rangle$, where angle brackets denote an average over space and infinite time. This total transport is fixed in our configuration; when heating exceeds cooling, excess heat is transported to the top boundary at an average rate that balances its production. This is seen by multiplying \eqref{eq: T eq} against $(1-z)$ and averaging to find $\langle \mathbf{\hat{z}}\cdot \vecu\hspace{1pt}T-\partial_zT\rangle =\langle (1-z)H\rangle$, where the right-hand average weights the local heating rate $H(z)$ by the distance $(1-z)$ that heat must travel to the top boundary. The mean \emph{conductive} transport $\langle-\partial_zT\rangle$, on the other hand, must be measured in simulations. Evaluating the vertical integral in $\langle-\partial_zT\rangle$ shows it is equal (in our dimensionless variables) to the mean temperature difference between the boundaries. The top temperature is zero by definition, so mean conductive transport is equal to the mean bottom temperature $\overline T(0)$, where an overline denotes an average over the horizontal directions and infinite time.

To reveal the analogies between transport in our configuration and in RBC we define a Nusselt number $Nu$ that is proportional to the ratio of mean total transport $\langle \mathbf{\hat{z}}\cdot \vecu\hspace{1pt}T-\partial_zT\rangle$ to mean conductive transport $\langle-\partial_zT\rangle$. Since total transport is fixed and conductive transport is equal to the mean bottom temperature $\overline T(0)$, normalizing $Nu$ to be unity in the static state $T_{st}(z)$ gives the definition $Nu=T_{st}(0)/\overline T(0)$. (The static state $T_{st}$ is the steady solution to the temperature equation with $\vecu=\mathbf 0$.) The diagnostic Rayleigh number $Ra$ described above, whose dimensional temperature scale is the mean temperature difference between the boundaries in the flow, can be written in terms of $Nu$ as $Ra=R/Nu$.

We simulate \eqref{eq: bouss} using a second-order energy-conserving finite difference code that has been benchmarked previously \cite{van2015pencil}. The control parameter $R$ is varied from $10^5$ to $10^{10}$ with $Pr=1$ fixed. The domain aspect ratio $\Gamma$ is chosen following Ref.~\cite{goluskin2016penetrative} to approximate large-domain values of $Nu$, for which the sufficiently large $\Gamma$ values are smaller when $R$ is larger. The spatial resolutions also follow Ref.~\cite{goluskin2016penetrative}, and we verify that the spatiotemporal integral relations $\left\langle {{\left| \nabla u \right|}^{2}} \right\rangle=R\left\langle \mathbf{\hat{z}}\cdot\vecu\hspace{1pt}T\right\rangle $ and $\left\langle {{\left| \nabla T \right|}^{2}} \right\rangle=\left\langle HT \right\rangle$, which follow from \eqref{eq: bouss} \citep{ostilla2015multiple,goluskin2016penetrative}, are satisfied to within $1\%$. The mesh is horizontally uniform, while in the vertical direction it has a clipped Chebyshev distribution so that points cluster in the boundary layers.

We have simulated IHC with exponential heating/cooling profiles of the form $H(z)=a e^{-z/\ell} - \beta$, as well as with uniform heating $H=1$, and for comparison we have simulated RBC in the same domains. In the dimensionless exponential profiles, $a=1/[\ell(1-e^{-1/\ell})]$ so that $\langle a e^{-z/\ell}\rangle=1$. The parameter $\beta$ is fixed to various values between $\beta=0$, at which there is nonuniform heating throughout the layer, and $\beta=1$, where there is net-zero heating/cooling since heating in the bottom region is equal to the cooling in the top region. The magnitude of the dimensionless profile $H(z)$ is partly arbitrary since the dynamics are equivalent if $H$ is scaled up by a factor while $R$ is scaled down by the same factor. These rescalings do not affect $Ra$, so to compare results across different heating profiles it is more meaningful if their $Ra$ values match rather than their $R$ values. The simulated $R$ range of $[10^5,10^{10}]$ corresponds to a different $Ra$ range for each different $H(z)$.
	

\begin{figure}[t]
	\center
	\includegraphics[width=0.48\textwidth,trim={10 12 10 13},clip]{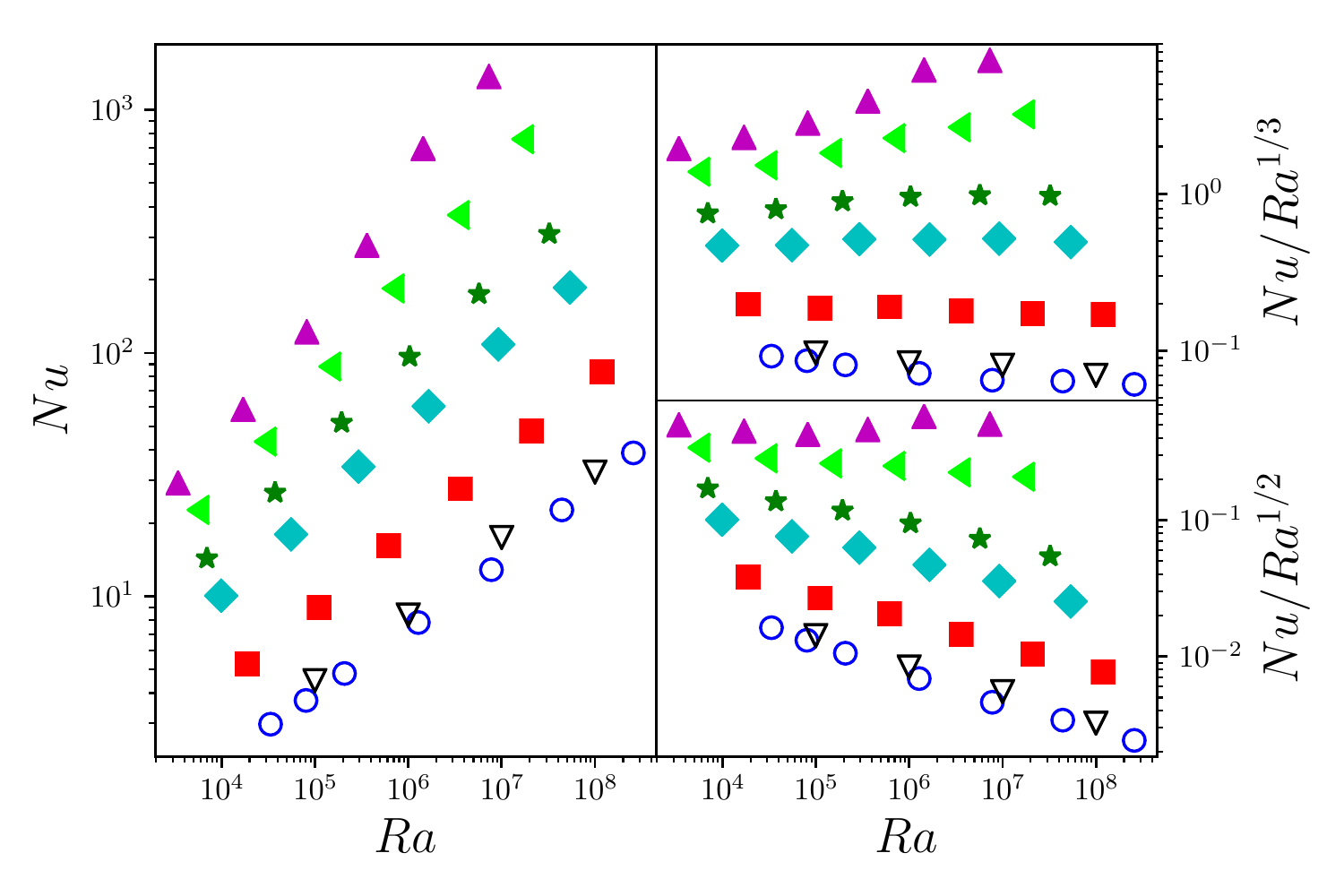}
	\caption{$Nu$ versus $Ra$ (left), and the same $Nu$ compensated by $Ra^{1/3}$ (top right) and by $Ra^{1/2}$ (bottom right). Cases shown are RBC \RB, uniform heating \IH, and non-uniform heating/cooling with exponential length scale $\ell=0.1$ and cooling strengths $\beta=0$ {\scriptsize\hlone}, $\beta=0.5$ \sfive, $\beta=0.7$ \sseven, $\beta=0.9$ {\scriptsize\snine}, and $\beta=1$ \sone.}
\label{fig:Nu}
\end{figure} 

Figure~\ref{fig:Nu} shows the dependence of $Nu$ on $Ra$ for IHC with various heating/cooling profiles, as well as for RBC. Three cases in the figure display best-fit scaling exponents that are less than 1/3, which is typical of the boundary-limited regime. The RBC simulations give $Nu\sim Ra^{0.29}$. In the IHC cases that are heated throughout, uniform heating gives $Nu\sim Ra^{0.29}$, and exponential heating without cooling ($\beta=0$) gives $Nu\sim Ra^{0.32}$. The latter scaling was found with exponential length scales of both $\ell=0.10$ and $\ell=0.05$, so we fix $\ell=0.10$ in further simulations where $\beta$ is varied, and the $\ell=0.05$ results are omitted from Fig.~\ref{fig:Nu} for clarity. The boundary-limited exponents are similar to those in past laboratory studies of IHC with an insulating bottom \cite[Table 3.1]{goluskin2016internally}. Such laboratory studies inevitably have nonuniform heating, but our results suggest that scaling exponents are not sensitive to the heating distribution as long as there is no cooling. To compare with past studies of IHC with equal top and bottom temperatures, one can define Nusselt and Rayleigh numbers using the mean temperature over the volume rather than over the bottom \cite{goluskin2016internally}. In these variables, 3D DNS with equal boundary temperatures \cite{goluskin2016penetrative} give an exponent of 0.26 (and 2D DNS are similar \cite{goluskin2012convection,goluskin2016penetrative,wang2020scaling}) while our simulations with uniform and exponential heating give the very close exponents of 0.28 and 0.27, respectively.

Whereas IHC with no cooling displays boundary-limited scaling, IHC with net-zero heating/cooling displays mixing-length scaling. In our simulations with $\beta=1$, the best-fit scaling is $Nu\sim Ra^{0.51}$. This is essentially the mixing-length exponent of 1/2, as was also found by Refs.~\cite{lepot2018radiative,bouillaut2019transition,miquel2020} with all boundaries insulating.

Having confirmed that heat transport displays boundary-limited or mixing-length scalings in the extreme cases where such behavior is expected based on prior studies, we turn to convection driven by unequal rates of heating and cooling. In particular, we have performed simulations with exponential heating/cooling profiles having $\ell=0.10$ and $\beta=0.5,0.7,0.9$, in addition to the $\beta=0$ and $\beta=1$ cases for which we have already reported boundary-limited and mixing-length scalings, respectively. At $\beta=0,0.5,0.7,0.9,1$, the corresponding heights below which $H(z)\ge0$ are $z_0=1,0.30,0.27,0.24,0.23$, and ratios of total cooling below $z_0$ to total heating above $z_0$ are $0,0.38,0.60,0.86,1$. The corresponding best-fit exponents of the $Nu\sim Ra^\gamma$ scalings are $\gamma=0.32,0.35,0.36,0.44,0.51$. These exponents span the range between the typical boundary-limited value of 1/3 and the mixing-length value of 1/2.

\begin{figure}[t]
\center
\includegraphics[width=0.48\textwidth,trim={0 0 0 1},clip]{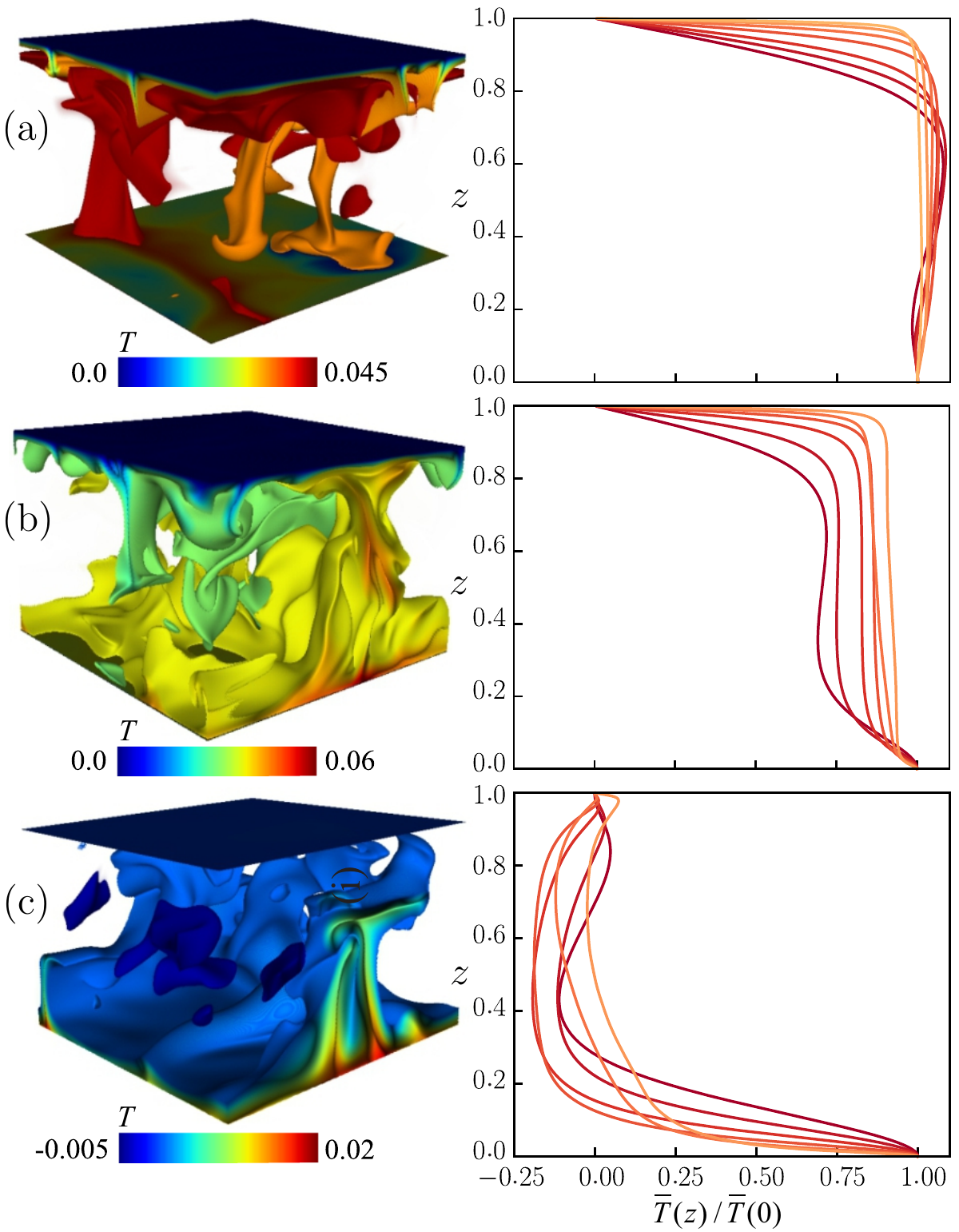}
\caption{Instantaneous temperature fields (left) at $R=10^8$ and normalized mean temperature profiles (right) at $R=10^5,10^6,10^7,10^8,10^9,10^{10}$ (dark to light) for (a) uniform heating, (b) exponentially distributed heating with $(\ell,\beta)=(0.1,0)$, and (c) net-zero exponential heating/cooling with $(\ell,\beta)=(0.1,1)$. In the instantaneous temperature visualizations, fluid is selectively made transparent to improve visualization.}
\label{fig:Tprofile}
\end{figure}

Turning to the structure of the simulated flows, Fig.~\ref{fig:Tprofile} shows instantaneous 3D temperature fields and mean temperature profiles $\overline T(z)$ for three of the extreme cases of IHC: heating that is (a) uniform or (b) exponentially distributed, and (c) net-zero heating/cooling. The temperature profiles are normalized by their bottom temperatures $\overline T(0)$. In case (a) the dominant temperature structures are falling cold plumes, and correspondingly $\overline T(z)$ shows an unstably stratified boundary layer at the top but none at the bottom. As $R$ is raised, increasingly strong turbulent mixing makes the temperature more uniform outside the boundary layer. In case (b), where the heating is positive everywhere but concentrated near the bottom boundary, there are not only falling cold plumes but rising hot plumes also. This is reflected in the $\overline T(z)$ profiles of Fig.~\ref{fig:Tprofile}(b), which have unstably stratified boundary layers at both boundaries. The temperature change across the bottom boundary layer shrinks, relative to the change across the top boundary layer, but the bottom boundary layer is still present at $R=10^{10}$. Nonetheless, despite the fact that the nonuniform heating in case (b) creates different flow structures than the uniform heating in case (a), the $Nu$--$Ra$ scaling is quite similar in both cases, as described above and reflected in Fig.~\ref{fig:Nu}. In the case of Fig.~\ref{fig:Tprofile}(c), where there is a net-zero exponential heating/cooling profile, the dominant temperature structures are rising hot plumes, and the corresponding $\overline T(z)$ profiles have unstably stratified thermal boundary layers at the bottom. Unlike in cases (a) and (b) of Fig.~\ref{fig:Tprofile}, the mean temperature profile $\overline T(z)$ remains far from isothermal in the turbulent interior, even with $R$ as large as $10^{10}$.

Much remains unknown about $R\to\infty$ asymptotic behavior of the configurations we have simulated. In IHC without cooling, which displays boundary-limited scalings in our simulations, it is unclear whether scaling exponents will increase towards their mixing-length values as the kinetic boundary layers become turbulent, as some believe occurs in RBC. In IHC with net-zero heating/cooling, we expect mixing-length scaling to persist as $R\to\infty$. This $R$-dependence would be consistent with the scaling arguments of Refs.\ \cite{bouillaut2019transition,miquel2020}, which additionally predict two different regimes of $Pr$-dependence. Two such regimes were indeed found with insulating boundaries \cite{bouillaut2019transition,miquel2020}, and we expect similar $Pr$-dependence with our isothermal top boundary, although we have not varied $Pr$ here. In IHC where unequal heating and cooling produce intermediate scaling exponents, as in our simulations with $\beta=0.7$ and $0.9$, there are thermal boundary layers at both the top and bottom boundaries (cf.\ Fig.\ 2 of the Supplement). The scaling of each boundary layer is needed to predict the scaling of the total temperature drop over the layer, which is the temperature difference used to define our Nusselt number. For the bottom boundary layer, the scaling arguments of Refs.\ \cite{bouillaut2019transition,miquel2020} apply and predict free-fall scaling for the temperature drop. This prediction merits future study, ideally in simulations where both $R$ and $Pr$ are varied. For the top boundary layer, the temperature drop is controlled by similar physics to RBC, so its asymptotic scaling is just as opaque as the asymptotic Nusselt number in RBC.

In summary, we have simulated convection in a fluid layer driven by various distributions of internal heating and cooling, as well as by heating alone. In configurations where all produced heat must cross a boundary, and so must traverse a boundary layer, the efficiency of heat transport is boundary-limited. This efficiency can be captured by a Nusselt number that is a ratio of total transport to conductive transport. In the boundary-limited regime, the dependence of this Nusselt number on a properly defined Rayleigh number has a scaling exponent close to $1/3$, as in the `classical' regime of RBC. In a configuration where none of the produced heat must cross a boundary because there is net-zero heating/cooling, we find a scaling exponent close to $1/2$, as in the conjectured `ultimate' regime of RBC. However, neither of these regimes capture convection driven by unequal heating and cooling in different regions, which is typical of various real-world systems such as planetary atmospheres. Our simulations suggest that unequal heating and cooling can, depending on their relative rates, produce any scaling exponent between about $1/3$ and~$1/2$

\noindent
\emph{Acknowledgements}\quad We are grateful to the Research Computing Data Core at the University of Houston for providing computational resources and technical support. DG was supported by NSERC Discovery Grant award RGPIN-2018-04263.
		
\bibliography{Ref.bib}

\end{document}